# Tailoring between network rigidity and nanosecond transient absorption in a-$Ge_xAs_{35-x}Se_{65}$ thin films

**Pritam Khan, Tarun Saxena and K. V. Adarsh***

[1]*Indian institute of Science Education & Research, Bhopal 462023, India*
**corresponding author: adarsh@iiserb.ac.in*

In this letter, we report the first observation of dramatic decrease in nanosecond (ns) pulsed laser induced transient absorption (TA) in a-$Ge_xAs_{35-x}Se_{65}$ thin films by tuning the amorphous network from floppy to rigid. Our results provide the direct experimental evidence of a self trapped exciton mechanism, where trapping of the excitons occurs through bond rearrangements. Taken together, a rigid amorphous network with more constraints than degrees of freedom, are unable to undergo any such bond rearrangements and results in weaker TA. However, we also demonstrate that excitation fluence can be effectively utilized as a simple tool to lift up enough constraints to introduce large TA even in rigid networks. Apart from this, we also show that TA is tunable with network rigidity as it blueshift when the mean coordination is increased from 2.35 to 2.6.

Amorphous chalcogenide thin films (ACTs) which contain group VI elements except oxygen as one or more of the major alloying components, are best known for their unique photosensitivity to bandgap or sub-bandgap light illumination [1, 2]. Owing to photosensitivity, they exhibit numerous photoinduced effects namely photodarkening (PD) [3], photobleaching [4, 5], photofluidity [6] etc. The physical mechanism behind such effects are believed to originate from the formation of self trapped excitons, also known as valence alternation pair (VAP) in localized band tail states having long excitation lifetime and strong electron-lattice coupling [7]. Such charged defects are relatively stable and allow structural rearrangement while decaying. Applications of photoinduced effects are well diverse ranging from waveguide designing [8], making micro lens [9], fabricating nano antenna [10] etc.

Among the different families of ACTs, in our present study, we have chosen Ge-As-Se films mainly because of two reasons: (1) its constituent components have similar size and electronegativity to form a close-to-ideal network [11] and (2) the system has a broad compositional range that forms amorphous network. Another advantage of this series of films is that they have good transparency in the infrared region and have optical nonlinearities as high as hundreds of times than that of silica glass [12]. Ge-As-Se glasses are prototype of covalently bonded amorphous solids and have been studied extensively to understand the role of topological constraints regarding the compositional variation of various physico-chemical properties. Further, such physico-chemical properties of these ACTs can be tuned by chemical composition which according to the constraint counting theory depend predominantly on the Mean Coordination Number (MCN) as denoted by <r>, where <r> is equal to the sum of the respective elemental concentrations times their covalent coordination number [13]. By calculating the angle constraints and bond length relative to the total number of degrees of freedom in an amorphous network by Philips et al. [14] predicted the existence of a percolation transition at $\langle r \rangle = 2.40$ from an under constrained "floppy" network to an over-constrained "rigid" one [14, 15]. Most previous studies with continuous wave (cw) illumination have shown that photostructural changes like PD, photorelaxation and photoexpansion disappears for <r>=2.4 [11, 16] while viscosity and heat capacity showed a global minimum at <r>=2.4 [17]. Nonetheless, such studies are completely missing on the optical response of intense short ns pulse illumination when the network is changed from floppy to rigid. Likewise, role of excitation fluence in determining the magnitude and kinetics of TA (induced absorption that persists only during illumination) is yet to be established. These studies are very important since they provide deeper insight into the self-trapped exciton mechanisms in their optical properties.

In this letter, we report 532 nm, 5ns short pulsed laser induced TA in a-$Ge_xAs_{35-x}Se_{65}$ thin films as a function of network rigidity. In a stark contradict to our previous results with cw illumination [18], we found that TA decreases dramatically with increasing MCN. Transient bond rearrangement through self trapped excitons is responsible for the large TA observed in the floppy system, whereas, rigid systems are unable to undergo such rearrangements and consequently show weaker effects. Interestingly, we showed that illumination with higher fluence could lift up the number of constraints locally to introduce large TA even in over coordinated glasses. Notably, TA spectra are tunable with network rigidity as it blue shift (to shorter wavelength) when MCN increases from 2.35 to 2.6.

The bulk samples of $Ge_xAs_{35-x}Se_{65}$ glasses with x=0, 5, 15, 20 and 25 were prepared by the melt quenching method starting with 99.999% pure Ge, As and Se powders. The cast samples were used as the source material for depositing thin films of average thickness ~1.0 μm by thermal evaporation in a vacuum of about $1\times10^{-6}$ mbar. A low deposition rate of 2-5A°/s was maintained to make the composition of the thin films close to that of the bulk. This was later confirmed with the EDAX measurements. The respective thin films and their MCNs are shown below.

Sample#1: $Ge_0As_{35}Se_{65}$: MCN=2.35, Sample#2: $Ge_5As_{30}Se_{65}$: MCN=2.4, Sample#3: $Ge_{15}As_{20}Se65$: MCN=2.5, Sample#4: $Ge_{20}As_{15}Se_{65}$: MCN=2.55 and Sample#5: $Ge_{25}As_{10}Se_{65}$: MCN=2.6. In the rest of the text, samples are identified with their MCN and composition. In our studies, we have not included any sample with MCN>2.6 since above this value multiple transitions are possible [19, 20] and also could manifest a chemical threshold for photo-oxidation of Ge [18].

For pump probe TA measurements, the pump beam was the second harmonics of the Nd:YAG laser (5ns pulses centered at 532 nm with an average fluence of 62 mJ/cm$^2$) used in single shot mode. The probe beam was selected from a Xenon Arc lamp (120 W) using a holographic grating with 1200 grooves/mm and delayed with respect to the pump beam using a digital delay generator. The pump and probe beams were overlapped at the sample, and the change in absorbance of the probe beam, $\Delta A = -\log[I_{es}/I_{gs}]$ at different delays was measured. Here $I_{es}$ and $I_{gs}$ are the transmitted intensities of probe beam after delay time t following the pump beam excitation and in ground state respectively. In our present study, except for the sample with MCN 2.35 and 2.4, thickness is coherent with the penetration depth of the pump beam laser. Therefore for those samples, the observed effects are mostly originating from surface.

Figure 1(a) shows the change in absorbance ($\Delta A$) of the samples at different MCN from 2.35 to 2.6 for the probe delay at 200 ns. From the figure it can be seen that TA (a measure of $\Delta A$) of all samples spreads over a broad wavelength range. By comparing the TA spectra we could observe two most important features:

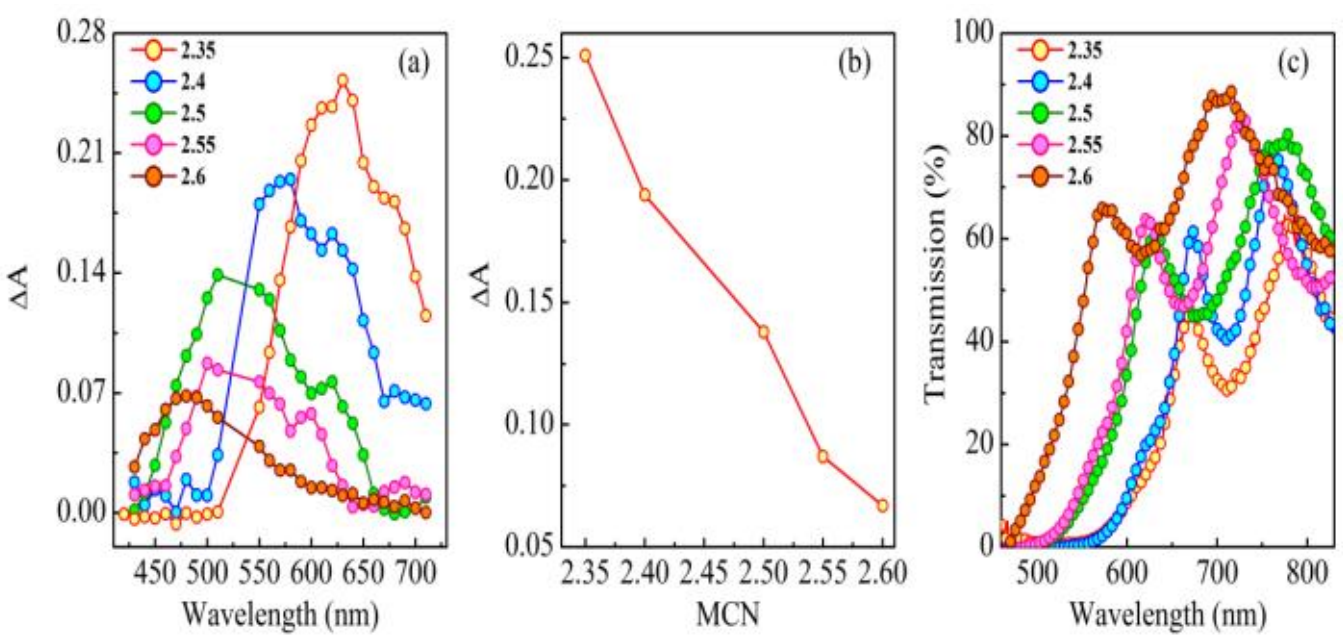

Figure 1 (a) TA spectra of a-$Ge_xSs_{35-x}Se_{65}$ thin films at a probe delay of 200ns. (b) Change in TA maxima with MCN. Evidently, TA decreases with increasing MCN. (C) Ground state transmission spectra of the same samples.

(1) Magnitude of TA decreases dramatically with increasing MCN (figure 1(b), in other words samples with floppy network show strong TA. Clearly this observation provides the first direct experimental evidence that ns light induced TA decreases with increasing MCN.

(2) TA spectra blue shift (absorption shifts to shorter wavelength) when the rigidity of the network increases as shown in figure 1(A). The observed blue shift strongly demonstrates that TA in a-$Ge_xAs_{35-x}Se_{65}$ thin films can be effectively tuned by network rigidity, which is otherwise not known previously.

In our next step to quantify the TA, we measured ground state transmission spectra of all samples and is shown in figure 1(d). Likewise TA, ground state transmission spectra also blue shift with increasing network rigidity. Such an observation essentially indicates that TA spectra are concurrent with ground state transmission spectra. In this regard, bandgap of all samples is calculated by classical Tauc plot and is found to be 1.75, 1.78, 1.83, 1.90 and 2.04 eV for MCN 2.35, 2.4, 2.5, 2.55 and 2.6 respectively. For the Tauc bandgap calculation, we have chosen the region of the transmission spectra where the value of absorption coefficient is more than $10^4$ cm$^{-1}$. From this analysis we can readily ascribe TA are due to interband transition. Naturally, the origins of such effects are believed to be instigated from the light induced structural changes, which is characteristic of ACTs.

Figure 2 (a) shows the change in TA maxima of the samples as a function of excitation fluence. As can be seen in the figure, TA maxima exhibits many fold increase with

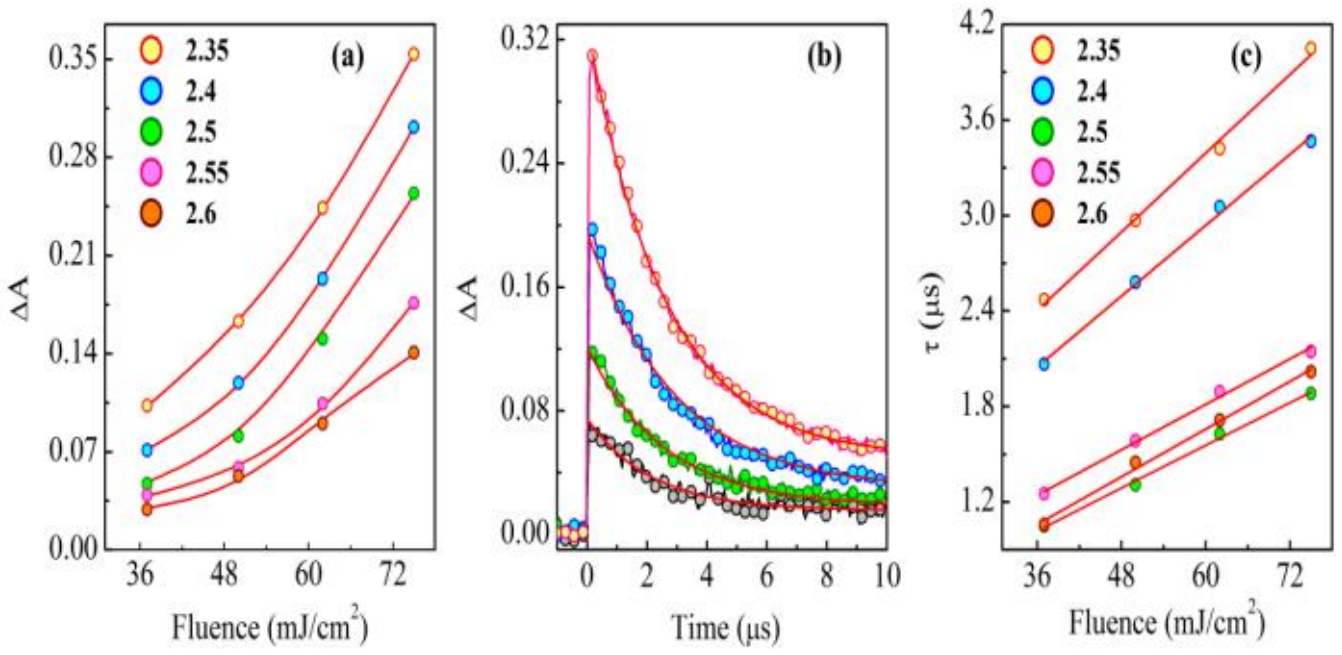

Figure 2 (a) Change in TA maximum for different MCN samples as a function of fluence. (b) Temporal evolution of TA maximum of a-$Ge_5As_{30}Se_{65}$ thin film at different excitation fluence. (c) Variation of decay time constant of TA at different excitation fluence which scales linearly. Colored symbols and the solid red line in each figure represent experimental data and theoretical fit respectively.

increasing fluence. The structural consistency of all the samples before and after illumination was confirmed with optical absorption (fig. 3(a) and AFM measurements (fig. 3(b) and (c)) which show that used fluency of pulsed laser does not lead to the ablation of the films.

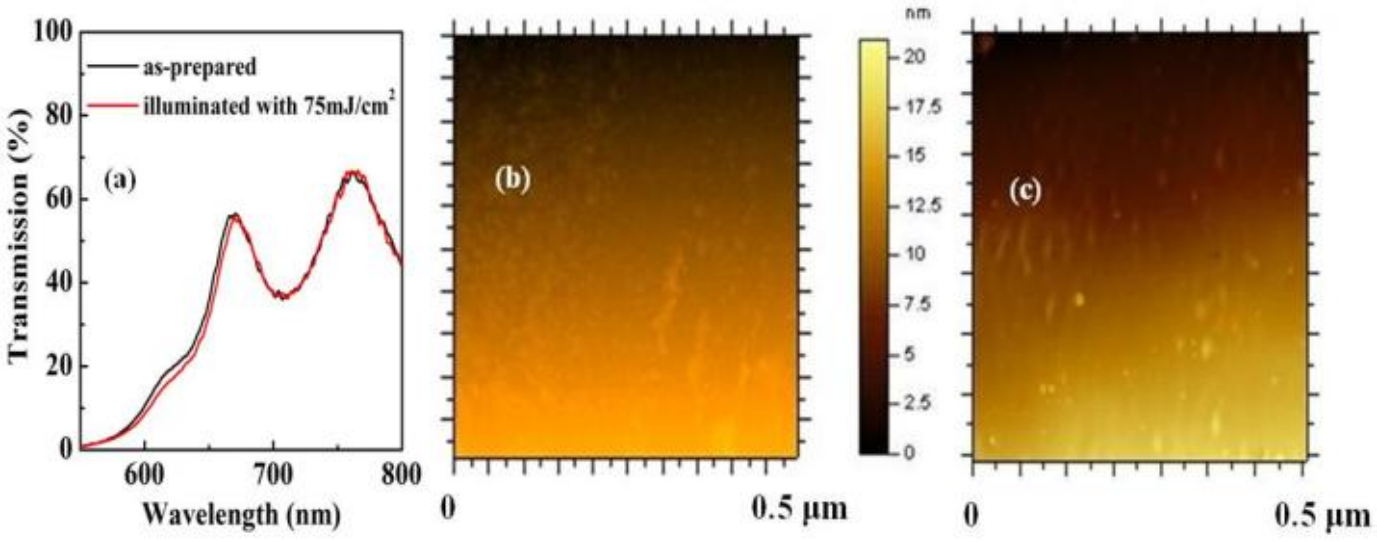

Figure 3 (a) Optical absorption spectra of as-prepared and illuminated (75mJ/cm$^2$) a-$Ge_5As_{30}Se_{65}$ thin films. (b) and (c) represents 2-d topographical image of the same in as-prepared and illuminated state (75mJ/cm$^2$).

Importantly, for all samples ΔA grows up non-linearly with excitation fluence and a second order polynomial equation fit very well to the experimental data. Such an observation suggests that the TA may be a two photon process, which can be understood from the double well model [2]. Following optical excitation, the local atomic configuration can relax either to the initial state (I) or to an intermediate state (F). The probability for the system to relax to F is much lower than to I in the case of single photon absorption. However, if two sites formed by the weakening or breaking of the inter atomic bonds are close enough, they form a complex B which can easily rearrange to F and is termed as two photon process [2]. Though each single photon has enough energy to produce a free electron-hole pair, but the probability to form state F is less. Nevertheless, at any excitation fluence, magnitude of TA is more in floppy than in rigid network. This result is expected because an over-constrained network has greater rigidity and as a result a high photon density is required to lift up a large number of structural constraints to introduce appreciable TA. However, at high excitation fluence even rigid samples show strong TA, which can be effectively utilized as a simple tool to introduce large TA even in rigid networks.

To understand the effect of excitation fluence on the kinetics of TA, we have plotted in figure 2(b) the temporal evolution of ΔA for MCN=2.4 at different laser fluence. Quite clearly, following the pump beam excitation, rise of TA is instantaneous which decays gradually. To get detailed information on the relaxation dynamics, we have performed the kinetics analysis at selected wavelengths (the wavelength at which TA shows maximum). For this we assumed that the excitation of ground state leads to population of a particular state. Naturally, change in population of this state can be determined by the following rate equation

$$\frac{dN_i}{dt} = -\sum_{j}^{n} a_{ij} N_i \qquad (1)$$

where $N_i$ and $a_{ij}$ ($i \neq j$) are population density of a state i and rate constants for transition from i to j states respectively. General solution of equ.1 is a linear combination of n exponentials and as a result, we have fitted our experimental data using the following equation

$$\Delta A = \sum_{j} A_i \exp - \left(\frac{t}{\tau_{ij}}\right) \qquad (2)$$

where ΔA, $A_i$, t, and $\tau_{ij}$ represent TA, amplitude of the exponential for a particular state, time and decay time constant for transition from i to j state respectively. Best fit to the experimental data clearly demonstrate that one decay constant is required to quantitatively model TA.

Figure 2(c) depicts the variation of τ calculated by using eq. (2) as a function of excitation fluence for all samples. Evidently, τ shows a linear dependence with a negative slope on excitation fluence, i.e. TA is found to decay faster at lower fluence. The slow decay of TA at higher fluence can be understood by considering the fact that permanent structural changes become prominent which makes the relaxation process drastically slower. This is further authenticated from figure 2(b) which shows that non-reversible component of TA (plateau of ΔA curve in the relaxation part) increases with increasing fluence, which instigates that the sample undergoes permanent changes.

The origin of ΔA and its concomitant decrease from floppy to rigid network can be well understood from non-radiative recombination of self trapped excitons [21]. Electron and hole in a solid are bound by an energy $e^2/4\pi\varepsilon\varepsilon_0 R$, where e is the electronic charge, ε and $\varepsilon_0$ are dielectric constant of the material and in air respectively. When illuminated with bandgap light, electrons and holes diffuse apart by a distance R which is quantified by the diffusion constant and thermalization time. There exists a critical distance $R_C$, known as the Coulomb capture (Onsager radius), defined by the following equation

$$e^2 \Big/ 4\pi\varepsilon\varepsilon_0 R_C = kT_0 \qquad (3)$$

where $T_0$ is the glass transition temperature. Now if $R > R_C$ they will diffuse further, however when $R < R_C$ condition is satisfied, electron hole will hop together to form an exciton (A) [21]. In our present study, for Ge-As-Se network glasses, ε is about 15 and $T_0$ is nearly 520K [22] and hence $R_C$ according to eq. (3) is ~ 20A°. On the otherhand, in the ground state electron hole pair is usually separated by a distance ~ 3A° [23]. Upon excitation, the separation increases by a factor $\sqrt{\beta t}$, where β is the diffusion coefficient and t is the thermalization time. For ACTs t is of the order of few picoseconds [24] and β= (1x10$^{-3}$cm$^2$/s) [25] that gives $\sqrt{\beta t}$ ~ 1.4A° which even after adding with initial separation is less than the critical distance $R_C$. Such situation strongly favors the excited electron-hole pairs to from excitons. The fraction of carriers form exciton is proportional to exp (-R/$R_c$).

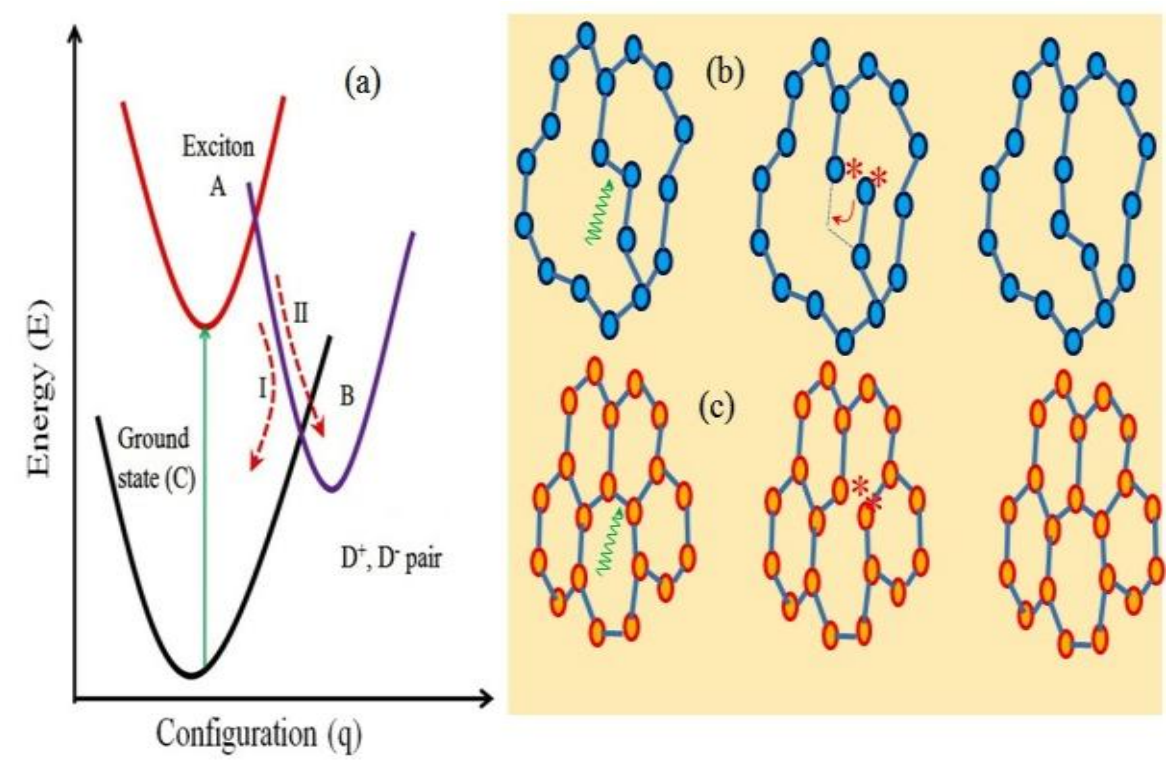

Figure 3 (a) Configurational Coordinate diagram showing two recombination channels of the excitons: (1) directly to ground state and (2) to a metastable state via the creation of defect pair. Lightinduced transient bonding rearrangement in (b) floppy and (c) rigid system.

These exciton recombination can take place via two different paths (1) direct return to ground state (C) or (2) via a metastable state (B) formed by $D^+$, $D^-$ pair (called as self trapped exciton), where D and the superscripts denote the defect and charge state respectively. Such decay of the excited states can be well understood from the configurational coordinate diagram [21] as shown in figure 4(a). As the ACTs posses very low photoluminescence efficiency, decay through path (1) is more probable than path (2). Here the formation energy ($E_d$) of $D^+$, $D^-$ pair is a small fraction of bandgap energy ($E_g$) and exciton energy ($E_x$) (corresponds to self-trapping of excitons) because of the strong electron-phonon coupling of the non-bonding lone pair orbital of chalcogen atom. Naturally the energy released in self trapping ($E_x$-$E_d$) is large fraction of $E_g$ which strongly accelerates non radiative recombination to the metastable state via self trapped exciton (path 2). The extra energy released is used to modify the structure of amorphous network through transient bond rearrangement, which gives rise to TA in the bandgap region. Consequently, the extent of TA is a measure of the strength of trapped excitons through transient bond rearrangement, which in turn strongly depend on the rigidity of the amorphous network and can be better understood from the schematic shown in figure 4(b) and (c). It can be seen from figure 3(b) that atoms in a floppy amorphous network can easily rearrange themselves between on-off states of the excitation. Upon excitation, each broken bond releases one constraint which temporally rearranges the network to produce large TA. On the otherhand, atoms in rigid systems are relatively constrained and their position remains mostly unaltered throughout the period of illumination [11, 16] as shown in figure 3(c). Although excitation can alleviate some local constraints, however, even then the numbers of constraints are more than the degrees of freedom. As a result, network cannot easily rearrange and the photoexcited state is therefore more likely to be in its original states with a little to no TA. Both of the above behavior can be connected to the topography of the energy landscape and to the rigidity of the samples. The strong TA in floppy network suggests that the light illumination permits the system to explore the surrounding minima on the energy landscape. Nevertheless, for a rigid system, energy landscape posses a few minima and therefore it can undergo a very little structural changes [16]. Needless to say, self-trapped excitons and the associated change in TA are found to be consistent with the network rigidity theory.

In conclusion, we have demonstrated that ns pulsed laser induced TA in a-$Ge_xAs_{35-x}Se_{65}$ thin films can be effectively tailored by tuning the MCN. This was evinced from a dramatic decrease in TA in our experiments, when the MCN was changed from 2.35 (floppy) to 2.60 (rigid). Light induced transient bonding rearrangement via self trapped exciton recombination mechanism accounts for the large TA in the floppy network. On the otherhand, rigid systems are unable to undergo any such bond rearrangements and consequently show weaker effects. Strikingly, TA of all samples shows a quadratic dependence on excitation fluence, whereas the kinetics (decay time constants) exhibits a linear relationship. Notably, TA spectra are compositionally tunable as it blue shifts with increase in MCN. Remarkably, our results provide new insights on synthesizing photostable and high damage threshold ACTs for intense pulsed laser applications.

The authors thank Department of Science and Technology (Project no: SR/S2/LOP-003/2010) and council of Scientific and Industrial Research, India, (grant No. 03(1250)/12/EMR-II) for financial support.